%% file: main.tex
\documentclass[orivec]{llncs}

\usepackage{latexsym,amssymb,amsmath}
\usepackage{url}
\usepackage[defblank]{paralist}
\usepackage{tikz}
\usetikzlibrary{arrows,shapes,shapes.multipart,snakes,automata,backgrounds,petri,positioning,shadows,matrix,decorations.pathmorphing,fit,positioning,calc,backgrounds}
\usepackage{todonotes}
\usepackage{xspace}
\usepackage{pifont}
\usepackage{listings}
\usepackage{parcolumns}
\usepackage{xcolor}
\usepackage{times}
\usepackage{lineno}
\usepackage{caption,subcaption}
\usepackage{algorithm}
\usepackage{algorithmic}

\input{macros-style}
\input{macros-common}
\input{macros}

\title{Soundness in Object-centric Workflow Petri Nets}
\author{Irina A. Lomazova$^1$ \and Alexey A. Mitsyuk$^1$ \and Andrey Rivkin$^2$}
\authorrunning{Lomazova, Mitsyuk, Rivkin}
\institute{%
$^1$HSE University (Russia)\\
$^2$Faculty of Computer Science, Free University of Bozen-Bolzano (Italy)
}

\begin{document}

\maketitle

\begin{abstract}
Recently introduced Petri net-based formalisms advocate the importance of proper representation and management of case objects as well as their co-evolution. In this work we build on top of one of such formalisms and introduce the notion of soundness for it. We demonstrate that for nets with non-deterministic synchronization between case objects, the soundness problem is decidable.
\end{abstract}

\input{0-introduction}

\input{1-motivating-example}
\input{1-preliminaries}
\input{2-formalism}

\input{3-wf-nets}

\input{4-soundness}

\input{5-related}

\input{6-conclusions}

\bibliographystyle{splncs04}
\bibliography{mybib}

\end{document}

%% file: macros-style.tex
\makeatletter
\g@addto@macro\normalsize{%
\setlength{\abovecaptionskip}{0pt}
\setlength{\belowcaptionskip}{-10pt}
\setlength\abovedisplayskip{3pt}
\setlength\belowdisplayskip{3pt}
\setlength\abovedisplayshortskip{3pt}
\setlength\belowdisplayshortskip{3pt}
}
\makeatother

%% file: macros-common.tex
\newcommand{\E}{\ensuremath{\mathcal{E}}}

\newcommand{\set}[1]{\{#1\}}                      

\newcommand{\tup}[1]{\langle #1\rangle}            

\makeatletter
\g@addto@macro\normalsize{%
\setlength{\abovecaptionskip}{-2pt}
\setlength{\belowcaptionskip}{12pt}
\setlength\abovedisplayskip{3pt}
\setlength\belowdisplayskip{3pt}
\setlength\abovedisplayshortskip{3pt}
\setlength\belowdisplayshortskip{3pt}
}
\makeatother

\newcounter{dummy} 
\newcounter{dummy1} 
\newcounter{dummy2}
\newcounter{dummy3} 
\newcounter{dummy4}
\newcounter{dummy5} 
\newcounter{dummy6}

\usepackage[thmmarks,amsmath]{ntheorem}
\theorempreskip{1pt}
\theorempostskip{1pt}

\theoremseparator{.}
\theorembodyfont{\itshape}
\theoremsymbol{$\triangleleft$}
\newtheorem{theorem}[dummy]{Theorem}
\newtheorem{lemma}[dummy1]{Lemma}
\newtheorem{definition}[dummy2]{Definition}
\newtheorem{proposition}[dummy3]{Proposition}

\theorembodyfont{\normalfont}
\newtheorem{example}[dummy4]{Example}
\newtheorem{remark}[dummy5]{Remark}

\theoremstyle{nonumberplain}
\theoremheaderfont{\itshape}
\theorembodyfont{\normalfont}

\theoremseparator{.}
\theoremsymbol{\ensuremath{\dashv}}
\newtheorem{proof}[dummy6]{Proof}

\qedsymbol{\ensuremath{\dashv}}


%% file: macros.tex
\newcommand{\mult}[1]{#1^\oplus}

\newcommand{\funsym}[1]{\mathtt{#1}}

\newcommand{\restr}[2]{\left.#1\right|_{#2}}

\newcommand{\typename}[1]{\mathtt{#1}}
\newcommand{\otypes}{\mathbb{O}}
\newcommand{\otype}{\typename{d}}

\newcommand{\naturals}{\mathbb{N}}

\newcommand{\onet}{OC-net\xspace}
\newcommand{\onets}{OC-nets\xspace}
\newcommand{\wfonet}{OC WF-net\xspace}
\newcommand{\wfonets}{OC WF-nets\xspace}

\newcommand{\p}{\mathrm{p}\xspace}

\newcommand{\net}{N}
\newcommand{\objnet}{ON}

\newcommand{\projection}[2]{\restr{#1}{#2}\xspace}
\newcommand{\typing}{\funsym{type}}
\newcommand{\places}{P}
\newcommand{\transitions}{T}
\newcommand{\flow}{F}
\newcommand{\enabled}[2]{#1[#2\rangle}
\newcommand{\fire}[3]{\enabled{#1}{#2}#3}
\newcommand{\marked}[2]{\tup{#1,#2}}

\newcommand{\reachable}[2]{\mathcal{R}(#1,#2)}

\newcommand{\marking}{\mathit{M}}
\newcommand{\markings}[1]{\mathbb{M}^{#1}}
\newcommand{\pre}[1]{{^\bullet{#1}}}
\newcommand{\post}[1]{{#1^\bullet}}

\newcommand{\inp}{\mathit{in}}
\newcommand{\outp}{\mathit{out}}

\newcommand{\tsys}[1]{\Gamma_{#1}}
\newcommand{\states}{S}

\definecolor{golden}{rgb}{1.0, 0.84, 0.0}
\definecolor{indigo}{rgb}{0.0, 0.25, 0.42}
\definecolor{mgreen}{rgb}{0.128,0.428,0}
\definecolor{burntorange}{rgb}{0.8, 0.33, 0.0}
\definecolor{camouflagegreen}{rgb}{0.47, 0.53, 0.42}
\definecolor{copperrose}{rgb}{0.6, 0.4, 0.4}

\tikzstyle{placelem}=[draw,
    cloud,
    cloud puffs = 15,
    minimum width=.2cm,
    minimum height=.2cm,
    fill=white,
    thick]

\tikzstyle{place}=[circle,thick,draw=black,fill=white,minimum size=7mm,font=\fontsize{9}{144}\selectfont]
\tikzstyle{transition}=[rectangle,thick,draw=black,fill=gray!20,minimum size=7mm]
\tikzstyle{enabledtransition}=[rectangle,very thick,draw=green!75,fill=green!20,minimum size=7mm]
 \tikzstyle{container}=[rectangle,rounded corners,very thick,draw=black!75,fill=black!20,minimum height=7mm,minimum width=14mm]
\tikzstyle{rplace}=[circle,ultra thick,draw=violet!75,fill=violet!20,minimum size=7mm]

\tikzstyle{erbox}=[draw, fill=gray!20, minimum width=7em, text width=6.0em, text centered,
  minimum height=3em,rounded corners,drop shadow]

\tikzstyle{placelem}=[draw,
    cloud,
    cloud puffs = 15,
    minimum width=.2cm,
    minimum height=.2cm,
    fill=white,
    thick]

\tikzstyle{netelem}=[draw,
    cloud,
    cloud puffs = 20,
    minimum width=1.8cm,
    minimum height=1.8cm,
    fill=blue!10,
    ultra thick]

\tikzstyle{relationselem}=[placelem,fill=pink!20]
\tikzstyle{noopelem}=[placelem,fill=orange!20]
\tikzstyle{enteredplace}=[place,fill=yellow!20]
\tikzstyle{boundplace}=[place,fill=yellow!30]
\tikzstyle{guardokplace}=[place,fill=yellow!40]
\tikzstyle{updatedplace}=[place,fill=yellow!50]
\tikzstyle{violplace}=[place,fill=red!10]
\tikzstyle{constrokplace}=[place,fill=green!10]
\tikzstyle{docommitplace}=[place,fill=green!30]
\tikzstyle{dorollbackplace}=[place,fill=red!30]
\tikzstyle{arc}=[-stealth',thick]
\tikzstyle{readarc}=[-,thick]

\tikzset{
state/.style={
       rectangle,
       fill=yellow!5,
       rounded corners,
       draw=black,  thick,
       minimum height=2em,
       minimum width=1.5cm,
       inner sep=1pt,
       text centered,
       }
}

%% file: 0-introduction.tex
\section{Introduction}

Traditional workflow nets often focus on a single case in isolation. 
However, in reality, the notion of a case is often more complex and may consist of multiple simultaneously ``active'' objects with complex inter-relations. 
This issue has been already recognized and addressed in various works. 
Among such, the seminal work on workflow patterns~\cite{AalstHKB03} with the related multiple instance patterns, 
works on proclets~\cite{AalstBEW00,AalstBEW01}, in which the authors studied interactions between multiple workflows.

When it comes to analyzing such nets, it is often assumed that, even in the presence of multiple, simultaneously active process instances each of which has its own lifecycle, the global system state is composed of their individual (local) states that, in turn, are rarely addressed in isolation. 
This becomes evident when analyzing the process correctness using the concept of soundness.
In the nutshell, the process is called sound if and only if any reachable process state, starting from its initial configuration, can properly terminate by reaching the final process configuration without leaving any intermediate process resources unused. 
Such soundness requires that the system model has clearly defined start and end points, and focuses only on the global termination.
This also implies that the model must have a clearly defined ``global'' lifecycle, forcing all the lifecycles it embeds to synchronize on the aforementioned global start and end points.

This assumption is quite common in practice, but it cannot capture a more complex setting in which, instead of the global lifecyle, one models an environment or a protocol without clearly defined global start and end points, and with process instances whose own lifecycles are treated as ``first class citizens'' (and thus their local termination comes to the fore).  
An example of such setting can be easily found in higher education, where process models can represent the whole educational lifecycle of a student of more refined scenarios such as work on a project for some course. 
Notice that such models can include multiple, possibly interacting participants (that is, active cases) and one might be interested in the proper completion only of one of them.
Like that the concept of the global state becomes redundant as each single case can be treated in isolation. 
All this calls for a new notion of soundness that is not based on global, terminal states. 

In this work, we address the aforementioned scenario by using the recently proposed formalism of object-centric Petri nets~\cite{AalstB20} -- a weak variant of colored Petri nets enriched with a special kind of transfer arcs. 
Using such nets, one can easily represent lifecycles of multiple objects and define their local start and end points. 
By drawing inspiration from multi-agent systems, where each agent is represented with a separate workflow and its successful completion in isolation can happen irrespectively from other workflows,  we introduce a new notion of soundness.
There, we focus on local terminal states only, assuming that a system can have any number of participants being simultaneously active. 
We propose constructive techniques for isolating behaviors of interest without alternating the global process behavior and introduce a new notion of soundness that focuses on an individual object lifecyle, treating all other objects as its environment. 
Moreover, we conjecture that checking this new variant of soundness is decidable.

%% file: 1-motivating-example.tex
\section{Motivating example}
\label{ex:students-projects}

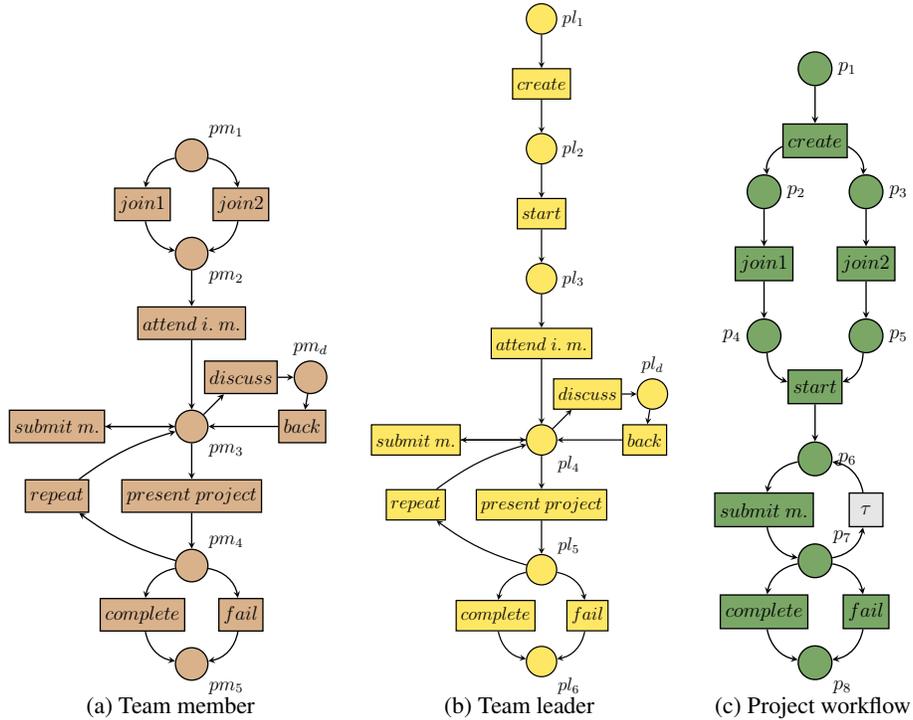
\begin{figure}[h!]
    \centering
    \begin{subfigure}[b]{0.36\textwidth}
        \centering
        \resizebox{.98\textwidth}{!}{\begin{tikzpicture}[->,>=stealth,double distance=1pt,auto,x=2mm,y=1.0cm,node distance=15mm and 15mm,thick, font=\large]
                
                \node[place, label = above right : $pm_1$, fill = brown!60] (pm1) {};
                \node[transition, below left of = pm1, fill = brown!60] (join1) {$join1$};
                \node[transition, below right of = pm1, fill = brown!60] (join2) {$join2$};
                \node[place, below right of = join1, label = below right : $pm_2$, fill = brown!60] (pm2) {};
                \node[transition, below of = pm2, fill = brown!60] (initmeet) {$attend\ i.\ m.$};
                \node[place, below= of initmeet, label = below right : $pm_3$, fill = brown!60] (pm3) {};
                \node[transition, left= of pm3, fill = brown!60] (submit) {$submit\ m.$};
                \node[transition, above right of = pm3, fill = brown!60] (discuss) {$discuss$};
                \node[transition, right = of pm3, fill = brown!60] (ediscuss) {$back$};
                \node[place, right of = discuss, label = above : $pm_d$, fill = brown!60] (pmd) {};
                \node[transition, below of = pm3, fill = brown!60] (finalmeet) {$present\ project$};
                \node[transition, below of = submit, fill = brown!60] (repeat) {$repeat$};
                \node[place, below of = finalmeet, label = above right : $pm_4$, fill = brown!60] (pm4) {};
                \node[transition, below left of = pm4, fill = brown!60] (complete) {$complete$};
                \node[transition, below right of = pm4, fill = brown!60] (fail) {$fail$};
                \node[place, below right of = complete, label = below right : $pm_5$, fill = brown!60] (pm5) {};
                
                \path[]
                (pm1) edge[bend right =35] (join1)
                (pm1) edge[bend left =35] (join2)
                (join1) edge[bend right =35] (pm2)
                (join2) edge[bend left =35] (pm2)
                (pm2) edge (initmeet)
                (initmeet) edge (pm3)
                (pm3) edge (submit)
                (pm3) edge (discuss)
                (submit) edge (pm3)
                (discuss) edge (pmd)
                (pmd) edge (ediscuss)
                (ediscuss) edge (pm3)
                (pm3) edge (finalmeet)
                (finalmeet) edge (pm4)
                (pm4) edge[bend right =35] (complete)
                (pm4) edge[bend left =35] (fail)
                (pm4) edge[bend left =10] (repeat)
                (repeat) edge[bend left =10] (pm3)
                (complete) edge[bend right =35] (pm5)
                (fail) edge[bend left =35] (pm5)
                ;
                
        \end{tikzpicture}}
        \caption{Team member}
        \label{fig:motivating-example-1}
    \end{subfigure}
    \hfill
    \begin{subfigure}[b]{0.36\textwidth}
        \centering
        \resizebox{.9\textwidth}{!}{\begin{tikzpicture}[->,>=stealth,double distance=1pt,auto,x=2mm,y=1.0cm,node distance=15mm and 15mm,thick, font=\large]
                
                \node[place, label = right : $pl_1$, fill = golden!60] (pl1) {};
                \node[transition, below of = pl1, fill = golden!60] (create) {$create$};
                \node[place, below of = create, label = right : $pl_2$, fill = golden!60] (pl2) {};
                \node[transition, below of = pl2, fill = golden!60] (start) {$start$};
                \node[place, below of = start, label = right : $pl_3$, fill = golden!60] (pl3) {};
                \node[transition, below of = pl3, fill = golden!60] (initmeet) {$attend\ i.\ m.$};
                \node[place, below= of initmeet, label = below right : $pl_4$, fill = golden!60] (pl4) {};
                \node[transition, left= of pl4, fill = golden!60] (submit) {$submit\ m.$};
                \node[transition, above right of = pl4, fill = golden!60] (discuss) {$discuss$};
                \node[transition, right = of pl4, fill = golden!60] (ediscuss) {$back$};
                \node[place, right of = discuss, label = above : $pl_d$, fill = golden!60] (pld) {};
                \node[transition, below of = pl4, fill = golden!60] (finalmeet) {$present\ project$};
                \node[place, below of = finalmeet, label = above right : $pl_5$, fill = golden!60] (pl5) {};
                \node[transition, below of = submit, fill = golden!60] (repeat) {$repeat$};
                \node[transition, below left of = pl5, fill = golden!60] (complete) {$complete$};
                \node[transition, below right of = pl5, fill = golden!60] (fail) {$fail$};
                \node[place, below right of = complete, label = below right : $pl_6$, fill = golden!60] (pl6) {};
                
                \path[]
                (pl1) edge (create)
                (create) edge (pl2)
                (pl2) edge (start)
                (start) edge (pl3)
                (pl3) edge (initmeet)
                (initmeet) edge (pl4)
                (pl4) edge (submit)
                (submit) edge (pl4)
                (pl4) edge (discuss)
                (discuss) edge (pld)
                (pld) edge (ediscuss)
                (ediscuss) edge (pl4)
                (pl4) edge (finalmeet)
                (finalmeet) edge (pl5)
                (pl5) edge[bend right =35] (complete)
                (pl5) edge[bend left =35] (fail)
                (pl5) edge[bend left =10] (repeat)
                (repeat) edge[bend left =10] (pl4)
                (complete) edge[bend right =35] (pl6)
                (fail) edge[bend left =35] (pl6)
                ;
                
        \end{tikzpicture}}
        \caption{Team leader}
        \label{fig:motivating-example-2}
    \end{subfigure}
    \hfill
    \begin{subfigure}[b]{0.24\textwidth}
        \centering
        \resizebox{.9\textwidth}{!}{\begin{tikzpicture}[->,>=stealth,double distance=1pt,auto,x=2mm,y=1.0cm,node distance=15mm and 15mm,thick, font=\large]
                
                \node[place, label = right : $p_1$, fill = mgreen!60] (p1) {};
                \node[transition, below of = p1, fill = mgreen!60] (create) {$create$};
                \node[place, below left of = create, label = right : $p_2$, fill = mgreen!60] (p2) {};
                \node[place, below right of = create, label = right : $p_3$, fill = mgreen!60] (p3) {};
                \node[transition, below of = p2, fill = mgreen!60] (join1) {$join1$};
                \node[transition, below of = p3, fill = mgreen!60] (join2) {$join2$};
                \node[place, below of = join1, label = left : $p_4$, fill = mgreen!60] (p4) {};
                \node[place, below of = join2, label = right : $p_5$, fill = mgreen!60] (p5) {};
                \node[transition, below right of = p4, fill = mgreen!60] (start) {$start$};
                \node[place, below of = start, label = right : $p_6$, fill = mgreen!60] (p6) {};
                \node[transition, below left of = p6, fill = mgreen!60] (submit) {$submit\ m.$};
                \node[transition, below right of = p6] (tau) {$\tau$};
                \node[place, below right of = submit, label = above right : $p_7$, fill = mgreen!60] (p7) {};
                \node[transition, below left of = p7, fill = mgreen!60] (complete) {$complete$};
                \node[transition, below right of = p7, fill = mgreen!60] (fail) {$fail$};
                \node[place, below right of = complete, label = below right : $p_8$, fill = mgreen!60] (p8) {};
                
                \path[]
                (p1) edge (create)
                (create) edge[bend right =35] (p2)
                (create) edge[bend left =35] (p3)
                (p2) edge (join1)
                (p3) edge (join2)
                (join1) edge (p4)
                (join2) edge (p5)
                (p4) edge[bend right =35] (start)
                (p5) edge[bend left =35] (start)
                (start) edge (p6)
                (p6) edge[bend right =35] (submit)
                (submit) edge[bend right =35] (p7)
                (p7) edge[bend right =35] (tau)
                (tau) edge[bend right =35] (p6)
                (p7) edge[bend right =35] (complete)
                (p7) edge[bend left =35] (fail)
                (complete) edge[bend right =35] (p8)
                (fail) edge[bend left =35] (p8)
                ;
                
        \end{tikzpicture}}
        \caption{Project workflow}
        \label{fig:motivating-example-3}
    \end{subfigure}
    \caption{Three different types of objects from the university project work scenario}
    \label{fig:motivating-example}
\end{figure}

We start by discussing an example that illustrates issues arising when one wants to model an environment consisting of multiple lifecycles and touches upon main ideas that we develop later on in our proposal. 
The example in Figure~\ref{fig:motivating-example} models a project work scenario in a university.
There, all three workflow nets model behaviors of three different entities (or objects): a project team leader, a project team member and a project.
Both project team leaders and members are students. 
Projects are performed in small teams each of which consists of three students: one team leader and two team members.

In Figure~\ref{fig:motivating-example}, separate entities are modeled using ordinary WF-nets.
Figure~\ref{fig:motivating-example-1} shows a workflow of a team member.
Figure~\ref{fig:motivating-example-2} illustrates how a team leader behaves.
Project itself has its own prescribed sequence of key milestones.
These make up a process workflow that is shown in Figure~\ref{fig:motivating-example-3}.
Let us consider behavior of separate entities in more detail.

We will begin from the project workflow shown in Figure~\ref{fig:motivating-example-3}.
Firstly, the project has to be $create$d.
It is done by a student who was chosen to be a team leader.
Student can start working on a project if there are three members in their team.
Thus, two team members have to $join$ to as soon as the team was created.
After this the students can start working on their project (this decision is taken by the team leader).
From project point of view, the team has to $submit$ part of the solution to supporting learning environment at least once.
There can be multiple submissions as well.
Then, the project finishes either by successful $complet$etion or by $fail$ure.

Let us consider now how student members behave.
One may notice similarities between models in Figure~\ref{fig:motivating-example-1} and Figure~\ref{fig:motivating-example-2}.
A team leader (see Figure~\ref{fig:motivating-example-2}) $create$s the project, then $start$s it.
After that each leader $attends$ the initial meeting and, at any time, can $discuss$ with other team members the project.
Any $submis$sion of each team need to be approved by the team leader.
The final point of each project is its $present$ation that can be attended by all the students participating to the project.
The project is evaluated based on the presentation, which may be found unsatisfactory.
In that case, the $present$ation can be $repeat$ed.
Otherwise, the project is be evaluated as completely $fail$ed or successfully $complet$ed.
A regular team member (see Figure~\ref{fig:motivating-example-1}) $join$s a team as first or second team member.
Then, they participate to the same project-related activities as team leaders do.

Notice these models are basic workflow nets.
In Figure~\ref{fig:motivating-example-1} and Figure~\ref{fig:motivating-example-2} black tokens in places represent a student's state, whereas in Figure~\ref{fig:motivating-example-3} tokens account for a project's state.

These three models show behavior of separate entities participating to the same scenario.
There are a lot of activities performed by several scenario participants together.
For example, activity $submit\ m.$ is performed in all three workflows: a student  submits the work, a team leader  approves this submission, a project changes its (progress) state.
Three students (leader and two members) always need to perform the initial meeting, whereas
$discuss$ions can be performed by two or more students.

\begin{figure}[t!]
    \centering
    \resizebox{.63\textwidth}{!}{\begin{tikzpicture}[->,>=stealth',auto,x=1cm,y=.8cm,node distance=20mm and 8mm,thick, font=\large]
            
            \node[place, label = above right : $pl_1$, fill = golden!60] (pl1) {};
            \node[transition, right of = pl1] (create) {$\textbf{create}$};
            \node[place, right of = create, label = above right : $p_1$, fill = mgreen!60] (p1) {};
            \node[place, below of = create, label = above right : $p_2$, fill = mgreen!60] (p2) {};
            \node[place, right of = p2, label = above right : $p_3$, fill = mgreen!60] (p3) {};
            \node[place, left of = p2, label = above left : $pm_1$, fill = brown!60] (pm1) {};
            \node[transition, below of = p2] (join1) {$\textbf{join1}$};
            \node[transition, below of = p3] (join2) {$\textbf{join2}$};
            \node[place, right of = join2, label = above left : $pl_2$, fill = golden!60] (pl2) {};
            \node[place, below of = join1, label = right : $p_4$, fill = mgreen!60] (p4) {};
            \node[place, below of = join2, label = right : $p_5$, fill = mgreen!60] (p5) {};
            \node[place, left of = p4, label = left : $pm_2$, fill = brown!60] (pm2) {};
            \node[transition, below of = p5] (start) {$\textbf{start}$};
            \node[place, left of = start, label = above : $pl_3$, fill = golden!60] (pl3) {};
            \node[transition, below of = pm2] (initmeeting) {$\textbf{attend\ i.\ m.}$};
            \node[transition, below of = initmeeting] (discuss) {$\textbf{discuss}$};
            \node[place, below left of = discuss, label = right : $pm_d$, fill = brown!60] (pmd) {};
            \node[place, below right of = discuss, label = left : $pl_d$, fill = golden!60] (pld) {};
            \node[place, left of = pmd, label = left : $pm_3$, fill = brown!60] (pm3) {};
            \node[place, right of = pld, label = right : $pl_4$, fill = golden!60] (pl4) {};
            \node[transition, below right of = pmd] (ediscuss) {$\textbf{back}$};
            \node[place, right of = pl4, label = right : $p_6$, fill = mgreen!60] (p6) {};
            \node[transition, below=20mm of pl4] (submit) {$\textbf{submit\ m.}$};
            \node[transition, below=20mm of pm3] (finalmeet) {$\textbf{present\ project}$};
            \node[transition, right of = submit] (tau1) {$\tau$};
            \node[place, below of = finalmeet, label = left : $pm_4$, fill = brown!60] (pm4) {};
            \node[place, right of = pm4, label = right : $pl_5$, fill = golden!60] (pl5) {};
            \node[place, below of = submit, label = right : $p_7$, fill = mgreen!60] (p7) {};
            \node[transition, below of = pl5] (complete) {$\textbf{complete}$};
            \node[transition, above right of = pl5] (repeat) {$\textbf{repeat}$};
            \node[transition, right of = complete] (fail) {$\textbf{fail}$};
            \node[place, below of = complete, label = right : $pl_6$, fill = golden!60] (pl6) {};
            \node[place, left of = pl6, label = left : $pm_5$, fill = brown!60] (pm5) {};
            \node[place, right of = pl6, label = right : $p_8$, fill = mgreen!60] (p8) {};

            \draw[double,->,brown!60] (pm3) -- (discuss);
            \draw[double,->,brown!60] (discuss) -- (pmd);
            \draw[double,->,brown!60] (pmd) -- (ediscuss);
            \draw[double,->,brown!60] (ediscuss) -- (pm3);
            
            \draw[double,->,golden!60] (pl4) -- (discuss);
            \draw[double,->,golden!60] (discuss) -- (pld);
            \draw[double,->,golden!60] (pld) -- (ediscuss);
            \draw[double,->,golden!60] (ediscuss) -- (pl4);
   
            \path[]
            (pl1) edge[golden!60] (create)
            (p1) edge[mgreen!60] (create)
            (create) edge[mgreen!60] (p2)
            (create) edge[mgreen!60] (p3)
            (create) edge[golden!60,bend left =35] (pl2)
            
            (pm1) edge[brown!60] (join1)
            (pm1) edge[brown!60] (join2)
            (p2) edge[mgreen!60] (join1)
            (p3) edge[mgreen!60] (join2)
            
            (join1) edge[mgreen!60] (p4)
            (join2) edge[mgreen!60] (p5)
            (join1) edge[brown!60] (pm2)
            (join2) edge[brown!60] (pm2)
            
            (pl2) edge[golden!60,bend left =25] (start)
            (p4) edge[mgreen!60] (start)
            (p5) edge[mgreen!60] (start)
            
            (start) edge[golden!60] (pl3)
            (start) edge[mgreen!60] (p6)
            
            (pl3) edge[golden!60] (initmeeting)
            (pm2) edge[brown!60] node {$2$} (initmeeting)
            
            (initmeeting) edge[brown!60] node {$2$} (pm3)
            (initmeeting) edge[golden!60] (pl4)
            
            (pl4) edge[golden!60,bend left =25] (finalmeet)
            (pm3) edge[brown!60] node {$2$} (finalmeet)
            
            (pm3) edge[brown!60,bend right =25] (submit)
            (submit) edge[brown!60,bend left =25] (pm3)
            (pl4) edge[golden!60] (submit)
            (submit) edge[golden!60] (pl4)
            
            (p6) edge[mgreen!60] (submit)
            (tau1) edge[mgreen!60] (p6)
            (submit) edge[mgreen!60] (p7)
            (p7) edge[mgreen!60] (tau1)
            
            (finalmeet) edge[brown!60] node {$2$} (pm4)
            (finalmeet) edge[golden!60] (pl5)

            (pm4) edge[brown!60] node {$2$} (complete)
            (pm4) edge[brown!60] node {$2$} (fail)
            (pm4) edge[brown!60] node {$2$} (repeat)
            (pl5) edge[golden!60] (complete)
            (pl5) edge[golden!60] (fail)
            (pl5) edge[golden!60] (repeat)
            (p7) edge[mgreen!60] (complete)
            (p7) edge[mgreen!60] (fail)
            
            (repeat) edge[brown!60,bend left =20] node {$2$} (pm3)
            (repeat) edge[golden!60,bend right =20] (pl4)
            
            (complete) edge[brown!60] node {$2$} (pm5)
            (fail) edge[brown!60] node {$2$} (pm5)
            (complete) edge[golden!60] (pl6)
            (fail) edge[golden!60] (pl6)
            (complete) edge[mgreen!60] (p8)
            (fail) edge[mgreen!60] (p8)

            ;
            
    \end{tikzpicture}}
    \vspace{.2cm}
    \caption{A Petri net model representing the university project scenario}
    \label{fig:motivating-example-merge}
\end{figure}
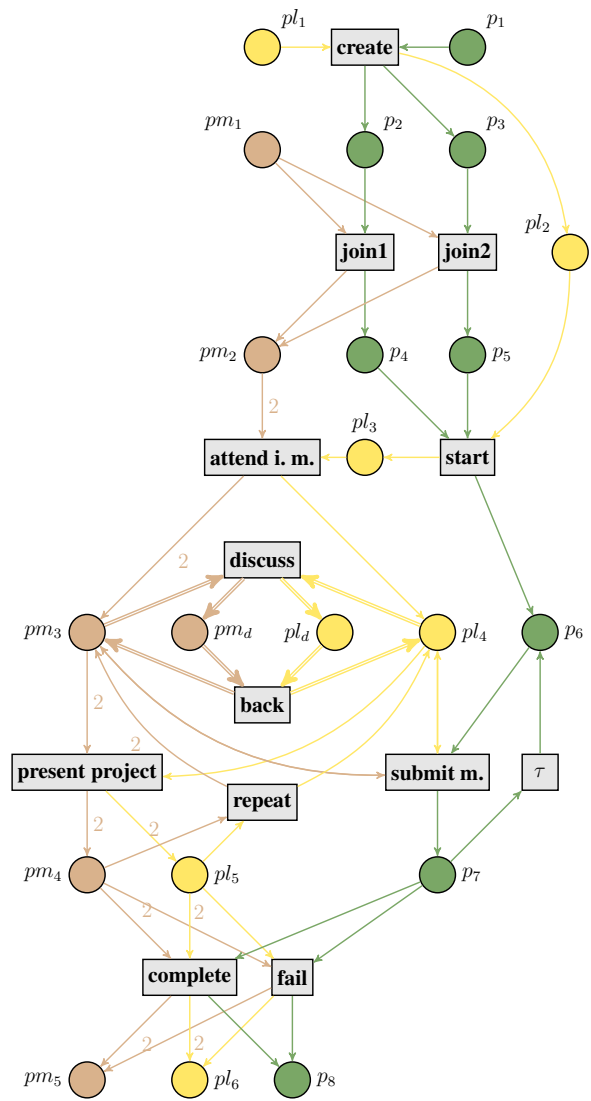
The question is how to model the whole process with all interactions between its participants.
This can be done using the formalism of object-centric Petri nets introduced in~\cite{AalstB20}.
Figure~\ref{fig:motivating-example-merge} shows how student and project workflows can be merged in the single object-centric worfklow net. 
To combine separate nets we merge transitions with same labels from all workflow nets.
It is important that we use one project net, one leader net, and two member nets.
This leads to weights on some of arcs in the merged \wfonet.
Moreover, $discuss$ions can be started and finished by an arbitrary number of students.
We use special \emph{variable} arcs\footnote{Upon transition firing, a variable arc transfers all the tokens from the input place it is connected to into the output place of the same type that is connected with another variable arc.} to model this fact.
Now we have a single model which represents the whole learning process with synchronous communications between process agents.
Note that asynchronous communications can always be modeled via synchronous mechanisms.

In this paper, we address the problem of \wfonet soundness.
It is known how to define and to check soundness of basic WF-nets.
For example, we can state that separate participants of our learning process from Figure~\ref{fig:motivating-example} are modeled by sound WF models.
But what can we say about soundness (correctness) of the \wfonet that is shown in Figure~\ref{fig:motivating-example-merge}?
And how soundness of agent models can be related to soundness of the \wfonet obtained by their merge?

%% file: 1-preliminaries.tex
\makeatletter
\newcommand{\xdasharrow}[2][->,>=angle 90]{
\tikz[baseline=-\the\dimexpr\fontdimen22\textfont2\relax]{
\node[anchor=south,font=\scriptsize, inner ysep=1.5pt,outer xsep=2.5pt](x){\ensuremath{#2}};
\draw[shorten <=3.4pt,shorten >=3.4pt,dashed,#1](x.south west)--(x.south east);
}
}
\section{Preliminaries}

\textbf{Multisets, Petri nets.} Let us start with fixing some standard notions related to \emph{multisets}. Given a finite set $B$,  a multiset $m$ over $B$ is the mapping of the form $m:B\rightarrow \naturals$.
Given an element $b \in B$, $m(b) \in \mathbb{N}$ denotes the number of times $b$ appears in the multiset. We write $b^n$ if $m(b)=n$. Given two multisets $m_1$ and $m_2$ over $B$:
\\
\begin{inparaenum}[\it (i)]
\item $m_1 \subseteq m_2$ (resp., $m_1 \subset m_2$) iff $m_1(b) \leq m_2(b)$ (resp., $m_1(b) < m_2(b)$) for each $b \in B$;\\
\item $(m_1 + m_2)(b) = m_1(b) + m_2(b)$;\\
\item if $m_1 \subseteq m_2$, $(m_2 - m_1)(b) = m_2(b) - m_1(b)$;\\
\item $|m|=\sum_{b\in B}m(b)$.
\end{inparaenum}
\\
We use $\mult B$ to denote the set of all finite multisets over $B$.

Given two disjoint sets $\places$ and $\transitions$ of places, multiset $\flow: (\places \times \transitions) \cup (\transitions \times \places) \rightarrow \naturals$ and function $\ell:\transitions\rightarrow A\cup\set{\tau}$, where $A$ is a finite set of activity labels and $\tau$ is a special symbol denoting silent transitions, we call $N=(\places,\transitions,\flow,\ell,A)$ a \emph{labeled place/transition net} (or a labeled P/T-net).

The notions of marking, transition enablement and firing  for classical P/T nets are defined as usual. We do not give precise definitions here, since we will not use them explicitly in what follows. 

\smallskip
\textbf{Transition systems, bisimulation equivalence.}
Given a finite set of action labels $A$ together with a special silent label $\tau$, a \emph{(labeled) transition system} (LTS) is a tuple $\tsys{}{} = (S,A,s_0, \to)$, where $S$ is a set of \emph{states}, $s_0$ is an \emph{initial state} and 
$\to \subset (S\times A \times S)$ is a \emph{transition relation}. 
In what follows, we shall write $s \xrightarrow{a} s'$ instead of $(s,a,s') \in \to$. 

\begin{definition}[Strong bisimulation]
\label{def:strong-bisimulation}
Let $\tsys{1} = (\states_1,A,s_{01},\to_1)$ and $\tsys{2} = (\states_2,A,s_{02},\to_2)$ be two labeled transition systems. Then relation $R\subseteq(\states_1 \times \states_2)$ is called a \emph{strong bisimulation} between $\tsys{1}$ and $\tsys{2}$ iff for every pair $(p,q)\in R$ and $a\in A$ the following holds:
\begin{enumerate}[\it (i)]
\item\label{def:strong-bisim-cond1} if $p\xrightarrow{a}_1 p'$, then there exists $q'\in\states_2$ such that $q \xrightarrow{a}_2 q'$ and $(p',q')\in R$;
\item\label{def:strong-bisim-cond2} if $q\xrightarrow{a}_2 q'$, then there exists $p'\in\states_1$ such that $p \xrightarrow{a}_1 p'$ and $(p',q')\in R$.
\end{enumerate}
\end{definition}

Given $a\in A$, $p \xdasharrow{~a~} q$ denotes a \emph{weak transition relation} that is defined as follows:
\begin{itemize}[$\bullet$]
\item $p \xdasharrow[->]{~a~} q$ iff $p (\xrightarrow{\tau})^* q_1\xrightarrow{a}q_2 (\xrightarrow{\tau})^* q$;
\item $p \xdasharrow[->]{\ensuremath{~\tau~}} q$ iff $p (\xrightarrow{\tau})^* q$.
\end{itemize}
Here, $(\xrightarrow{\tau})^*$ denotes the reflexive and transitive closure of $\xrightarrow{\tau}$.

\begin{definition}[Weak bisimulation]
\label{def:weak-bisimulation}
Let $\tsys{1} = (\states_1,A,s_{01},\to_1)$ and $\tsys{2} = (\states_2,A,s_{02},\to_2)$ be two labeled transition systems. Then relation $R\subseteq(\states_1 \times \states_2)$ is called a \emph{weak bisimulation} between $\tsys{1}$ and $\tsys{2}$ iff for every pair $(p,q)\in R$ and $a\in A$ the following holds:
\begin{enumerate}[\it (i)]
\item\label{def:weak-bisim-cond1} if $p\xrightarrow{a}_1 p'$, then there exists $q'\in\states_2$ such that $q \xdasharrow{~a~}_2 q'$ and $(p',q')\in R$;
\item\label{def:weak-bisim-cond2} if $q\xrightarrow{a}_2 q'$, then there exists $p'\in\states_1$ such that $p \xdasharrow{~a~}_1 p'$ and $(p',q')\in R$.
\end{enumerate}
\end{definition}
We say that a state $p\in\states_1$ is strongly (weakly) bisimilar to $q \in \states_2$, written $p \sim  q$ (correspondingly, $p \approx  q$), if there exists 
a strong (weak) bisimulation $R$ between $\tsys{1}$ and $\tsys{2}$ such that $(p,q)\in R$. 
Finally, $\tsys{1}$ is said to be strongly (weakly) bisimilar to $\tsys{2}$ , written $\tsys{1}  \sim \tsys{2}$ (correspondingly, $\tsys{1}  \approx \tsys{2}$), if $s_{01}  \sim s_{02}$ ($s_{01}  \approx s_{02}$).

\medskip

A \emph{Workflow net} is a labeled P/T net with two special places: $\inp$ and $\outp$. These places are used to mark the beginning and the ending of the workflow process. 

\begin{definition}[Workflow net]

A labeled P/T net $N=(\places,\transitions,\flow,\ell,A)$ is called a \emph{workflow net} (\emph{WF-net}) iff:
\begin{enumerate}
\item  
There is one source place $\inp\in \places$ and one sink place $\outp\in \places$, s. t. $\pre{\inp}=\post{\outp}=\emptyset$.
\item
Every node from $\places \cup \transitions$ is on a path from $\inp$ to $\outp$.
\item  
The initial marking in $N$ contains the only token in its source place and is denoted by $[\inp]$.
\end{enumerate}
\end{definition}

The marking with the only token in the sink place $\outp$ is called \emph{final} and is denoted by $[\outp]$. 

%% file: 2-formalism.tex
\section{Object-centric Petri nets}

In this section, we formally define the class of object-aware workflow Petri nets, called \emph{Object-Centric Petri nets (OC-nets)}. Specifically, we partially adopt the formalism of object Petri nets presented in~\cite{AalstB20}, which was introduced as a target model for multi-perspective process mining. 

OC-nets can be thought of as a rather limited version of colored Petri nets (CPNs), where each color type contains exactly one color. Thus, the color types coincide with the colors, tokens in the place labeled with the color type are indistinguishable from each other, in which case we do not need variables in arc expressions, but  only the natural numbers denoting the multiplicity of the arc, as in P/T-nets. 

By analogy with~\cite{AalstB20}, colors in OC-nets are called \emph{object types}.  We denote the set of object types as $\otypes$, and reserve $\E$ to denote a special ``empty'' type.
As in~\cite{AalstB20}, OC-nets also allow to use \emph{variable arcs} --- a special type of  arcs for simultaneous transferring  of a non-deterministic amount of tokens from one place to another.

\begin{definition}[Object-centric net]
\label{def:onet}
An \onet $\objnet$ is a tuple  
$(\otypes, \places,\transitions,\flow,\typing,\ell,A)$, where:
\begin{enumerate}
\item $\places$ and $\transitions$ are finite sets of places and transitions, s.t. $P\cap T=\emptyset$;
\item $\typing: \places \rightarrow \otypes$ is a place typing function;
\item $\flow: E  \rightarrow \naturals\cup\set{\mu}$ is a multiset of flow arcs, where $E=(\places \times \transitions) \cup (\transitions \times \places)$ and $\mu$ is used to indicate \emph{variable} arcs;
\item for every $p\in\places$ and $t\in\transitions$, 
	if $\flow(t,p)=\mu$, then 
	\begin{itemize}
	\item there is \emph{exactly one} $p^\prime\in\places$ s.t. $\flow(p^\prime,t)=\mu$ and $\typing(p)=\typing(p^\prime)$, and
	\item there is no $p^{\prime\prime}\in\places$ s.t. $\typing(p)=\typing(p^{\prime\prime})$, $\flow(p^{\prime\prime},t)>0$ or $\flow(t,p^{\prime\prime})>0$;
	\end{itemize}
\item $A$ is a finite set of activity names, and $\ell:\transitions\to A\cup\set{\tau}$ is a function which maps transitions to activities with $\tau$ denoting a silent (invisible) activity.
\end{enumerate}
\end{definition}

The meaning of variable arcs in OC-nets remains in line with~\cite{AalstB20}: they are meant to model the transfer of multiple objects from one place to another. 
The  condition 4 in the above definition is similar to the consistent variability condition from~\cite{AalstB20} stipulating that for any transition $t$ and any object type $d$ there can be no more than one pair of ingoing and outgoing for $t$ variable  arcs adjacent to places of type $d$.
For ease of notation, we denote with $\transitions_\mu$ the set of all transitions $t\in\transitions$ s.t. $\flow(t,p)=\mu$, for some $p\in\places$. 

\medskip

Since the tokens residing in the same place cannot be distinguished, the OC-net marking is defined similarly to the marking for P/T-nets. More specifically, a \emph{marking} of \onet $\objnet = (\otypes, \places,\transitions,\flow,\typing)$ is a function $\marking:\places\rightarrow \naturals$. 
When $\marking(p)=n$ and $n>0$ for some $p\in\places$, we say that there are $n$ objects of type $\typing(p)$ in state $M$.
We write $\marked{\objnet}{\marking}$ to denote \onet $\objnet$ marked with $\marking$ and use symbol $\marking_0$ to define the initial marking of the net. 
For ease of notation, we also denote with $[p_1^{i_1},\ldots,\p_n^{i_n}]$ a concrete multiset representing a marking in which each place $p_k$ contains $i_k$ tokens. Finally, with $\markings{\places}$ we denote the countably infinite set of all possible markings defined on top of $\places$.

\medskip
Let us now specify the net dynamics. 
As customary, given $x\in\places\cup\transitions$, we use $\pre{x}:=\set{x\mid \flow(\_,x)>0 \text{ or }\flow(\_,x)=\mu}$ to denote the \emph{preset} of $x$ and  $\post{x}:=\set{x\mid \flow(x,\_)>0 \text{ or }\flow(x,\_)=\mu}$ to denote the \emph{postset} of $x$. 

Since a transition can have multiple incoming and outgoing variable arcs of different types, we would like to be able to explicitly identify how many tokens from each place adjacent to the variable arc are transferred by the transition firing. To this end, we introduce a function $\alpha:\otypes\to\naturals$ specifying the transition's \emph{transfer mode}. Note that by Definition~\ref{def:onet}, a transition $t$ in an \onet cannot have two transitions $p_1,p_2\in\pre{t}$ s.t. $\flow(p_1,t)=\flow(p_2,t)=\mu$. Thus, $\alpha$ does not create ambiguities by relating to object types only.

We then say that transition $t$ is \emph{enabled} in marking $M$ with transfer mode $\alpha$, written $\enabled{\marking}{t,\alpha}$, iff, for every $p\in\pre{t}$, the following conditions hold:
\begin{enumerate}[\it (i)]
\item $\flow(p,t)\leq \marking(p)$, if $\flow(p,t)\in\naturals$, and 
\item $\marking(p)\geq\alpha(\typing(p))$, if $\flow(p,t)=\mu$.
\end{enumerate}
When $t$ is enabled in marking $M$ with transfer mode $\alpha$, it may \emph{fire}, 
yielding new marking $\marking'$ that is defined for every $p\in\places$ as follows:
\[\marking'(p)=
		\begin{cases}  
		\marking(p)-\alpha(\typing(p)), & \text{if \,}\flow(p,t)=\mu\\
		\marking(p)+\alpha(\typing(p)), & \text{if \,}\flow(t,p)=\mu\\
		\marking(p)-\flow(p,t)+\flow(t,p), & \text{otherwise}
		\end{cases}\]
We denote this as $\fire{\marking}{t,\alpha}{\marking^\prime}$ and assume that the definition is inductively extended to sequences $\sigma\in(\transitions\times \otypes\to \naturals)^*$ of transition firings. We say that $\marking'$ is \emph{reachable} from $\marking$ if there exists $\sigma\in(\transitions\times \otypes\to \naturals)^*$, s.t. $\fire{\marking}{\sigma}{\marking^\prime}$.  For an \onet $\objnet$, we write $\reachable{\objnet}{\marking_0}$ to denote the set of all markings reachable from its initial marking $\marking_0$. 

The execution semantics of an \onet can be captured with a possibly infinite-state LTS that accounts for all possible executions starting from {\onet}'s initial marking. Formally, an \onet $\objnet=(\otypes, \places,\transitions,\flow,\typing,\ell,A)$ with initial marking $M_0$ \emph{induces} a labeled transition system $\tsys{\objnet} = (S,A,s_0, \to)$, where: 
\begin{itemize}
	 \item $S=\reachable{\objnet}{\marking_0}$ and $s_0=M_0$,
	 \item for $M,M^\prime\in S$ it holds that: 
	 	$M\xrightarrow{a} M^\prime$ iff  $\fire{\marking}{t,\alpha}{\marking^\prime}$,  for some $t\in\transitions$ s.t. $\ell(t)=a$  and some transfer mode $\alpha$. 
\end{itemize}

\begin{algorithm}[t!]
\caption{Variable arc elimination}
\begin{algorithmic} \label{alg:arc-elim}
\REQUIRE $\tup{ON, M_0}$, where $\objnet=(\otypes, \places,\transitions,\flow,\typing,\ell,A)$, s.t. for some $t\in\transitions$ and $p\in\places$ it holds that $\flow(t,p)=\mu$
\ENSURE $\tup{ON^\prime, M_0^\prime}$, where $\objnet^\prime=(\otypes^\prime, \places^\prime,\transitions^\prime,\flow^\prime,\typing^\prime,\ell^\prime,A)$, s.t. for all $t\in\transitions^\prime$ and $p\in\places^\prime$ it holds that $\flow^\prime(t,p)\neq\mu$
	\STATE $\otypes^\prime=\otypes \cup \set{\E}$ 
	\STATE $\places^\prime=\places\cup\set{lock}$
	\STATE $\transitions^\prime = T$
	\FORALL{$t\in \transitions_\mu$} 
			\STATE $\places^\prime=\places^\prime\cup \set{lock_t}$
			\STATE $\transitions^\prime = \transitions^\prime \cup \set{start_t,add_t}$
	\ENDFOR
	\FORALL{$p,p^\prime\in \places^\prime, t\in \transitions^\prime$} 
		\STATE $\flow^\prime(p,t)=
		\begin{cases}  
		k, & \text{if \,}\flow(p,t)=k\\
		1, & \text{if \,}(p,t)=(lock,t) \\
		1, & \text{if \,}\flow(p,t)=\flow(t,p^\prime)=\mu \,\AND\, (p,t)\in\set{(p,start_t),(lock,start_t),\\
		&\phantom{xxxxxxxxxxxxxxxxxxxxxxxxxx}\,(lock_t,add_t), (p,add_t), \\
		&\phantom{xxxxxxxxxxxxxxxxxxxxxxxxxx}\,(lock_t,t), (p^\prime,t)}
		\end{cases}$
		\STATE $\flow^\prime(t,p^\prime)=
		\begin{cases}  
		k, & \text{if \,}\flow(t,p^\prime)=k\\
		1, & \text{if \,}(p,t)=(t,lock) \\
		1, & \text{if \,}\flow(p,t)=\flow(t,p^\prime)=\mu \,\AND\, (p,t)\in\set{(start_t,p), (start_t,lock_t),\\
		&\phantom{xxxxxxxxxxxxxxxxxxxxxxxxxx}\,(add_t,lock_t),(add_t,p^\prime), (t,p^\prime)}
		\end{cases}$
		\STATE $\typing^\prime(p)=
		\begin{cases}
		\typing(p), & \text{if \,}p\in\places\\
		\E, & \text{otherwise}
		\end{cases}$
		 ~~~~~~~   $\ell^\prime=
		\begin{cases}
		\ell(t), & \text{if \,}t\in\transitions\\
		\tau, & \text{otherwise}
		\end{cases}$
		\STATE $M_0^\prime=
		\begin{cases}
		M_0(p), & \text{if \,}p\in\places\\
		1, & \text{if \,}p = lock\\
		0, & \text{otherwise}
		\end{cases}$
	\ENDFOR
	
\end{algorithmic}
\end{algorithm}

Now, after we have defined the syntax and semantics of \onets, we would like to observe that every \onet $\objnet$ can be represented with \onet $\objnet'$ that does not contain variable arcs. To this end, we provide a variable arc elimination algorithm and show that its output is always weakly bisimilar to its input.
The weak bisimilarity is conditioned by the fact that in order to correctly represent the behavior of $\objnet$, $\objnet'$ must include additional intermediate steps that are, however, invisible and therefore not relevant to the comparison of net behavior. That is why we need to use a form of bisimulation allowing to ``skip'' transitions irrelevant for the behavioral comparison~\cite{Milner89}. 
Finally, we say that two marked \onets $\tup{\objnet,M_0}$ and $\tup{\objnet^\prime,M^\prime_0}$ are \emph{weakly bisimilar} (and denote it as $\tup{\objnet,M_0}\approx \tup{\objnet^\prime,M^\prime_0}$) if for transition systems $\tsys{\objnet}$ and $\tsys{\objnet^\prime}$ they respectively induce it holds that $\tsys{\objnet}\approx \tsys{\objnet^\prime}$.

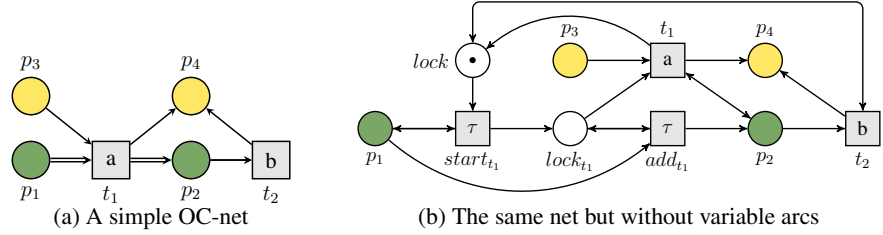
\begin{figure}[t!]
\centering
\begin{subfigure}[b]{0.38\textwidth}
         \centering
			\resizebox{.8\textwidth}{!}{\begin{tikzpicture}[->,>=stealth,double distance=1pt,auto,x=2mm,y=1.0cm,node distance=15mm and 15mm,thick, font=\large]
				\node[transition,  label = below : $t_1$] (t1) {a};			
				\node[place, fill = mgreen!60, left of = t1, label = below : $p_1$] (p1) {};				
  				\node[place,  fill = mgreen!60, right of = t1, label = below : $p_2$] (p2) {};
				\node[place, fill = golden!60, left of = t1, yshift=1.2cm, label = above : $p_3$] (p3) {};		
				\node[place, fill = golden!60, right of = t1, yshift=1.2cm, label = above : $p_4$] (p4) {};	
				\node[transition, right of = p2,  label = below : $t_2$] (t2) {b};			

				\draw[double,->] (p1) -- (t1);    
    			\draw[double,->] (t1) -- (p2);   
    			\draw[->] (p2) -- (t2);    
    			\draw[->] (p3) -- (t1);    
    			\draw[->] (t1) -- (p4);    
    			\draw[->] (p2) -- (t2);			
				\draw[->] (t2) -- (p4);
  			\end{tikzpicture}}
         \caption{A simple \onet }

         \label{fig:obj_net-to-wf_net-1}
     \end{subfigure}
     \hfill
     \begin{subfigure}[b]{0.6\textwidth}
         \centering
			\resizebox{.95\textwidth}{!}{\begin{tikzpicture}[->,>=stealth',auto,x=1cm,y=.8cm,node distance=20mm and 8mm,thick, font=\large]
			
			\node[place, fill = mgreen!60, label = below : $p_1$] (p1) {};				
			\node[transition, right of = p1, label = below : $start_{t_1}$] (start1) {$\tau$};
			\node[place,  right of = start1, label = below : $lock_{t_1}$] (lock) {};
			\node[transition, right of = lock, label = below : $add_{t_1}$] (add) {$\tau$};
			\node[place, fill = mgreen!60, right of = add, label = below : $p_2$] (p2) {};
			\node[transition, above of = add, yshift=-6mm, label = above : $t_1$] (end) {a};
			\node[place, fill = golden!60, left of = end, label = above : $p_3$] (p3) {};
			\node[place, fill = golden!60, right of = end, label = above : $p_4$] (p4) {};
			\node[place, above of = start1, yshift=-6mm, label = left : $lock$] (glock) {$\Large\bullet$};		
			\node[transition, right of = p2, label = below : $t_2$] (t2) {b};

			\path[]
			 	(p1) edge (start1)    
			 	(start1) edge (p1)
			 	(start1) edge (lock)
			 	(lock) edge (add)
			 	(add) edge (lock)
				(add) edge (p2)
				(p1) edge[bend right =40] (add)
  				(lock) edge (end)
  				(glock) edge (start1)
  				(end) edge[bend right=40] (glock)
  				(p3) edge (end)
  				(end) edge (p4)
  				(p2) edge (t2)
  				(t2) edge (p4)
    			;   
    		\draw[<->] (end) edge (p2);
    		\draw[<->, rounded corners] (glock)  -- ++ (0,1.5) -|  (t2);

  			\end{tikzpicture}}
  		\caption{The same net but without variable arcs}
         \label{fig:obj_net-to-wf_net-2}
     \end{subfigure}
     \caption{Demonstration of Algorithm~\ref{alg:arc-elim} applied to  \onet~\ref{fig:obj_net-to-wf_net-1} with  \onet~\ref{fig:obj_net-to-wf_net-2} as output}
     \label{fig:obj_net-to-wf_net}
\end{figure}

Given a generic \onet, Algorithm~\ref{alg:arc-elim} allows to construct from it a  behaviorally equivalent \onet that has no variable arcs. 
For example, consider a simple net in Figure~\ref{fig:obj_net-to-wf_net-1}. By applying the above algorithm to it, we obtain a new \onet (see Figure~\ref{fig:obj_net-to-wf_net-2}) that has no variable arcs. 
As it has been mentioned before, the algorithm produces a net that behaves exactly like the input one.
Indeed, the net in Figure~\ref{fig:obj_net-to-wf_net-2} models removed variable arcs with a sub-net representing a ``lossy'' transfer, 
working under the assumption that at least one token gets transferred from $\inp$ to $\outp$ (this assumption is in line with Definition~\ref{def:onet}). 
First, using $start_t$, the net enters into a critical section that is guarded by the special lock place $lock$. 
Notice that this lock is global to the whole net and is used to guard firing of every transition from the original net.
When the net acquires the lock, it can ``transfer'' tokens from $p_1$ to $p_2$ using transition $add_{t_1}$. 
As soon as at least one token has been generated in $p_2$, the transfer can be finalized by firing $t_1$ that releases the global lock.
This example leads us to the following statement.

\begin{proposition}\label{prop:bisimilarity}
Let $\tup{\objnet, M_0}$ be a marked \onet, where $\objnet=(\otypes, \places,\transitions,\flow,\typing,\ell,A)$. 
Then we can effectively construct a marked \onet $\tup{\objnet^\prime, M_0^\prime}$, such that $\objnet^\prime$ has no variable arcs and $\tup{\objnet,M_0}\approx \tup{\objnet^\prime,M^\prime_0}$.
\end{proposition}

\bigskip
	
Recall that 
\onets specify a class of nets that can be used to model the behavior of multiple, possibly interacting, objects of different types  within one system. Notice that the formalism of \onets does not allow you to distinguish between objects of the same type. 
However, at the same time, one can focus on the lifecycle represented by one type of objects. Unlike the life-cycle of a single object, the lifecycle of an object type can encompass various possible behaviours  for objects of that type.
Next, we introduce a definition of a \emph{projection} that supports retrieving a  subnet based on a given object type. 

 Let $\objnet=(\otypes, \places,\transitions,\flow,\typing,\ell,A)$ be an \onet, $M$ be a marking in $\objnet$, $d\in \otypes$. In what follows, we will make use of the following notations for \emph{$\otype$-based restrictions} of \onet components:  
	 $\restr{\places}{\otype}=\set{p\mid p\in\places,\,\typing(p)=\otype}$,
	  $\restr{\transitions}{\otype}=\set{t\mid t\in\transitions,\,\text{ there is }p\in\pre{t}\cup\post{t}\text{ s.t. }\typing(p)=\otype}$,
	  $\restr{\flow}{\otype}= \flow \cap 
		((\restr{\places}{\otype}\times \restr{\transitions}{\otype})\cup(\restr{\transitions}{\otype}\times\restr{\places}{\otype}))$
	 $\restr{\typing}{\otype}=\restr{\typing}{{\restr{\places}{\otype}}}$, 
	 $\restr{\ell}{\otype}=\restr{\ell}{{\restr{\transitions}{\otype}}}$,
	and 
	$\restr{M}{\otype}=\restr{M}{{\restr{\places}{\otype}}}$. 
	
	\begin{definition}[Object type projection]
\label{def:projection}
A \emph{$\otype$-typed projection} of an \onet $\objnet=(\otypes, \places,\transitions,\flow,\typing,\ell,A)$ is the \onet $(\set{\otype}, \restr{\places}{\otype},\restr{\transitions}{\otype},\restr{\flow}{\otype},\restr{\typing}{\otype},\restr{\ell}{\otype},A)$. We denote the projection as $\projection{\objnet}{\otype}$.
\end{definition}
	
	Since all places in an object-type projection of an OC-net are labeled with the same and only type, this labeling can be omitted, and the object projection can be considered as a P/T-net. Then, abusing the notation, we denote  both the $\otype$-typed projection of  \onet $\objnet$ and the corresponding P/T-net by $\projection{\objnet}{\otype}$.
	\medskip
	
Finally, let $d' \not\in \otypes$ be a 'fresh' name for a data type. By $\objnet_{[\otype/\otype']}$ we denote the \onet $\objnet'=(\otypes', \places',\transitions',\flow',\typing',\ell',A)$ such that:
\begin{itemize}
	\item graphs $(\places,\transitions,\flow)$ and $(\places',\transitions',\flow')$ are isomorphic via some isomorphism function $\iota$, 
	\item $\otypes'=(\otypes\setminus\set{\otype})\cup\set{\otype'}$,
	\item  for $p\in \places$, $\typing'(\iota(p)) = \otype'$ if $\typing(p) = \otype$, otherwise  $\typing'(\iota(p)) = \typing(p)$,
	\item for $t \in \transitions$, $\ell'(\iota(t))= \ell(t)$.
\end{itemize}
In other words, $\objnet_{[\otype/\otype^\prime]}$ is a copy of $\objnet$ in which $\otype$ is replaced with $\otype'$.

%% file: 3-wf-nets.tex
\section{Object-centric Workflows}

In this section, we show how object-centric nets can be used  to represent workflows of multiple, possibly interacting, artifacts of different types.

A classical, untyped, workflow net 
is used to represent workflows or lifecycles of one single object (or case) in isolation. There are also generalized workflow nets \cite{HeeSV04}, where the initial state contains several indistinguishable objects (black dots) in the source place, and the final state contains all these objects in the sink place, so that a workflow run terminates when all  items initially residing in the source place are processed and located in the sink place.

Next, we will show how to extend the concept of a workflow net to cover  multiple objects of different types. For this we use OC-nets, in which a typed place is a location for tokens of one type. In OC workflow nets, tokens of a specific type will be processed according to a specific workflow, but can interact with each other and with tokens of other types. Another important feature is that new objects can appear in source places at any moment. Thus, while it is assumed that  the proper behavior of each object  terminates, the system represented by the OC workflow net can run infinitely long.

\begin{definition}
\label{def:wfonet}
An \onet $\objnet=(\otypes, \places,\transitions,\flow,\typing,\ell,A)$ is called an \emph{object-centric worfklow net} (or \wfonet for short), iff:
\begin{enumerate}
\item for every $\otype\in\otypes$, $\projection{\objnet}{\otype}$ is a WF-net;
\item for every transition $t \in \transitions$, $\pre{t}\cup \post{t} \not = \emptyset$, i.e. there are no isolated transitions in $\objnet$.
\end{enumerate}
\end{definition}

Thus, a classical WF-net is the \wfonet with exactly one object type, and \wfonets form true extension of WF-nets. 

For \wfonet $\objnet=(\otypes, \places,\transitions,\flow,\typing,\ell,A)$, we use the notation  $\inp_\otype\in\places$ and $\outp_\otype\in\places$ for its (only) \emph{source} and \emph{sink} places of type $d$, i.e. $\typing(\inp_\otype) = \typing(\outp_\otype) = \otype$ and $\pre{\inp_\otype}=\post{\outp_\otype}=\emptyset$. By $\places_\inp$ and $\places_\outp$, we denote the sets of all source and sink places, respectively.

For \wfonets, we allow several tokens in each of its source places for an initial marking, not restricting the consideration to a workflow for just one case and just one initial marking. We also do not 
suppose that there are some fixed final markings for \wfonets and that reaching a final marking means termination of the system. 
\medskip

\begin{figure}[t!]
\centering
\begin{subfigure}[b]{0.47\textwidth}
         \centering
			\resizebox{.8\textwidth}{!}{\begin{tikzpicture}[->,>=stealth',auto,x=2mm,y=1.0cm,node distance=15mm and 15mm,thick, font=\large]
				\node[place, fill = golden!60, label = below : $\inp_1$] (in1) {};				
				\node[transition, right of = in1, label = below : $a$] (a1) {};
  				\node[place, fill = golden!60, right of = a1, label = below : $p_1$] (p1) {};
				\node[transition, right of = p1, label = below : $b$] (b1) {};
				\node[place, fill = golden!60, right of = b1, label = below : $\outp_1$] (out1) {};
			
				\node[place, fill = mgreen!60, below of = in1, yshift =-5mm, label = below : $\inp_2$] (in2) {};				
				\node[transition, right of = in2, label = below : $b$] (b2) {};
  				\node[place, fill = mgreen!60, right of = b2, label = below : $p_2$] (p2) {};
				\node[transition, right of = p2, label = below : $a$] (a2) {};
				\node[place, fill = mgreen!60, right of = a2, label = below : $\outp_2$] (out2) {};				
				
				\path[]
   				 	(in1) edge (a1)    
    				(a1) edge (p1)
    				(p1) edge (b1)
    				(b1) edge (out1)
    				(in2) edge (b2)    
    				(b2) edge (p2)
    				(p2) edge (a2)
    				(a2) edge (out2)
    				;   
    		
  			\end{tikzpicture}}
         \caption{}

         \label{fig:net-ex11}
     \end{subfigure}
     \hfill
     \begin{subfigure}[b]{0.47\textwidth}
         \centering
			\resizebox{.8\textwidth}{!}{\begin{tikzpicture}[->,>=stealth',auto,x=2mm,y=1.0cm,node distance=15mm and 15mm,thick, font=\large]
				\node[place,fill = golden!60, label = below : $\inp_1$] (in1) {};				
				\node[transition, right of = in1, label = below : $a$] (a) {};
  				\node[place, fill = golden!60, right of = a1, label = below : $p_1$] (p1) {};
				\node[transition, right of = p1, label = below : $b$] (b) {};
				\node[place, fill = golden!60, right of = b1, label = below : $\outp_1$] (out1) {};
				\node[place, fill = mgreen!60, below of = in1, label = below : $\outp_2$] (out2) {};				
  				\node[place, fill = mgreen!60, below of = p1, label = below : $\inp_2$] (in2) {};
				\node[place, fill = mgreen!60, above of = p1, label = below : $p_2$] (p2) {};				
				
				\path[]
   				 	(in1) edge (a)    
    				(a) edge (p1)
    				(p1) edge (b)
    				(b) edge (out1)
    				(a) edge (out2)    
    				(in2) edge (b)
    				;   		
    			
    			\draw[rounded corners=2mm] (b) |- (p2);
    			\draw[rounded corners=2mm] (p2) -| (a);
    			
  			\end{tikzpicture}}
  		\caption{}
         \label{fig:net-ex12}
     \end{subfigure}
     \caption{Two simple WF-nets (Figure~\ref{fig:net-ex11}) and their synchronous composition (Figure~\ref{fig:net-ex12})}
     \label{fig:net-ex1}
\end{figure}
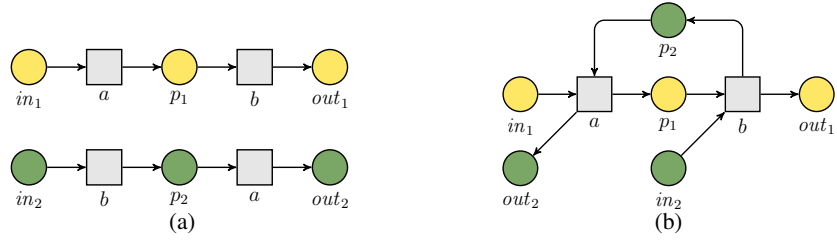

 The \emph{synchronous composition} of two nets $N_1$ and $N_2$ is a net obtained by pairwise merging of transitions in $N_1$ with  transitions with the same labels in $N_2$. Fig.~\ref{fig:net-ex1} gives an example of a synchronous composition of two nets. Note that though both components in 
Fig.~\ref{fig:net-ex11} are simple and sound WF-nets, their synchronous composition is  dead in its initial marking, provided  the initial marking is the “canonical” one (that is, only input places have tokens assigned to them).

It is easy to note that every \wfonet $\objnet=(\otypes, \places,\transitions,\flow,\typing,\ell,A)$ is a synchronous composition of all its $\otype$-projections  $\projection{N}{\otype}$, where  $\otype\in\otypes$, and
 $\otype$-projection  $\projection{N}{\otype}$ represents the lifecycle for objects of  type $\otypes$.

So far, we have been considering nets allowing to model co-evolution of multiple objects of various object types. 
The main limitation of the \onet formalism is that the identity of such objects is always hidden: one can only see how many anonymous objects of the same type are stored in a place.
Yet, in certain scenarios it might be still useful to \emph{track} a single object. 
This tracking can be achieved by, first, isolating the behavior of the object and then carefully merging the extracted subnet with the original net. 
To this end, we first define a new, more restrictive type of projection that, instead of focusing on all possible behaviors of a single object type, captures behavior of one object.

\begin{definition}[1-safe projection]
\label{def:projection}
Let $\objnet=(\otypes, \places,\transitions,\flow,\typing,\ell,A)$ be an \onet, $\otype \in \otypes$. \onet $\objnet'$ is called  \emph{1-safe $\otype$-typed projection} of  \onet $\objnet$ (denoted $\projection{\objnet}{\otype}^1$) iff $\objnet'$ coincides with $\projection{\objnet}{\otype}$ except the arc multiplicity function, which takes the value 1 throughout its domain. Namely, $\projection{\objnet}{\otype}^1 = (\set{\otype}, \restr{\places}{\otype},\restr{\transitions}{\otype},\restr{\flow}{\otype}^1,\restr{\typing}{\otype},\restr{\ell}{\otype},A)$, where $\restr{\flow}{\otype}^1(x,y) =1$, if $\restr{\flow}{\otype}(x,y) \not = 0$, and $\restr{\flow}{\otype}^1(x,y) = 0$, otherwise.
\end{definition}

Now, using  1-safe projection, we can construct a net that can  track  a single object evolution as well as its potential interactions with other objects of the same or any other type in the net.

\begin{definition}[Tracking \wfonet]
\label{def:projection}
Let $\objnet=(\otypes, \places,\transitions,\flow,\typing,\ell,A)$ be a \wfonet and $\objnet'=(\set{\otype'}, \places',\transitions', \flow',\typing',\ell',A)$, such that $\objnet'=(\projection{\objnet}{\otype}^1)_{[\otype/\otype']}$.
Then its \emph{$(\otype,\otype')$-typed tracking extension}, denoted as $\objnet_{\otype+\otype'}$, is the \wfonet $(\otypes\cup\set{\otype'}, \places \cup \places',\transitions\cup\transitions'\cup\transitions'',\flow'',\typing\cup\typing',\ell'',A)$\footnote{Here, given two functions $f:X\to A$ and $g:Y\to B$, we write $f\cup g:X\cup Y\to A\cup B$ iff $\restr{f}{X\cap Y}=\restr{g}{X\cap Y}$.}, where:
\begin{itemize}
	\item $\transitions''= \set{t''\mid t\in T_1 \cup T_2}$, where $T_1 = (\bigcup_{p\in\places'\cap\places}\pre{p})\setminus\transitions'$ and $T_2 = \transitions'\cap\transitions_\mu$;
	\item $F^{\prime\prime}(x,y)=\begin{cases}   
	F(x,y), & \text{if  \,} (x,y)\in(\places\times\transitions)\cup(\transitions\times\places)\\
	F^\prime(x,y), & \text{if \,} (x,y)\in(\places^\prime\times\transitions^\prime)\cup(\transitions^\prime\times\places^\prime)\\
	F(x,t) - 1, & \text{if \,} (x,y)\in(\places\times\transitions^{\prime})\text{, }t\in\transitions\\ 
	F(t,x) - 1, & \text{if \,} (x,y)\in(\transitions^{\prime}\times\places)\text{, }t\in\transitions\\ 
	F(x,t), & \text{if \,} (x,y)\in(\places\times (T_1\cup T_2))\text{, }t\in\transitions\\
	F(t,x), & \text{if \,} (x,y)\in((T_1\cup T_2)\times\places)\text{, }t\in\transitions\\
	\end{cases}$
\end{itemize}
\end{definition}

For $(\otype,\otype')$-typed tracking extension of \wfonet $\objnet$, we call $\objnet$ the \emph{ground component}, and $\objnet'=(\projection{\objnet}{\otype}^1)_{[\otype/\otype']}$ -- the \emph{tracking component}.

\begin{proposition} 
Let $\objnet=(\otypes, \places,\transitions,\flow,\typing,\ell,A)$ be a \wfonet, $M_\inp$ -- its initial marking. Let also $\otype\in \otypes$ and  $\otype'\not \in \otypes$. 

Then $\objnet$ with the initial marking $M_\inp$ is weakly bisimilar to its $(\otype, \otype')$-typed tracking extension $\objnet_{\otype+\otype'}$ with the initial marking  $M_{\inp'}$, s.t.  

$M_{\inp'}(\inp_{\otype'})=1$, if $M_\inp(\inp_{\otype})\geq 1$, $M_{\inp'}(\inp_{\otype'})=0$, if $M_\inp(\inp_{\otype})=0$,

$M_{\inp'}(\inp_{\otype})= M_{\inp'}(\inp_{\otype})-1$, if $M_\inp(\inp_{\otype})\geq 1$, $M_{\inp'}(\inp_{\otype})=0$, if $M_\inp(\inp_{\otype})=0$,

for every $p\in \places \setminus \{\inp_{\otype}\}$, $M_{\inp'}(p)=M_{\inp}(p)$.
\end{proposition}

Notice that the tracking extension itself can be used for monitoring concrete objects.
The property below demonstrates that one can essentially monitor a finite number of objects at once. 

\begin{proposition}
Let $\objnet=(\otypes, \places,\transitions,\flow,\typing,\ell,A)$ be a \wfonet. Let also $d_1, d_2\in \otypes$,  $d_1', d_2'\not \in \otypes$, and $d_1\not = d_2$, $d_1'\not = d_2'$. 

Then nets $(\objnet_{\otype_1+\otype_1'})_{\otype_2+\otype_2'}$ and 
$(\objnet_{\otype_2+\otype_2'})_{\otype_1+\otype_1'}$ are identical, i.e. the result of the successive application of multiple tracking extensions does not depend of their order.
\end{proposition}

\begin{example}
\label{ex:tracking-OCWF-net}
\input{2-1-example}
\end{example}

%% file: 2-1-example.tex
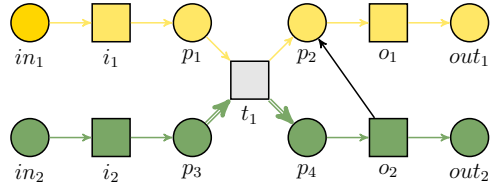
\begin{figure}[t!]
    \centering
    \resizebox{.53\textwidth}{!}{\begin{tikzpicture}[->,>=stealth',auto,x=1cm,y=.8cm,node distance=15mm and 5mm,thick, font=\large]
            
            \node[place, label = below : $in_1$, fill = golden] (in1) {};
            \node[transition, right of = in1, label = below : $i_1$, fill = golden!60] (i1) {};
            \node[place, right of = i1, label = below : $p_1$, fill = golden!60] (p1) {};
            \node[transition, below right of = p1, label = below : $t_1$] (t1) {};
            \node[place, below left of = t1, label = below : $p_3$, fill = mgreen!60] (p3) {};
            \node[transition, left of = p3, label = below : $i_2$, fill = mgreen!60] (i2) {};
            \node[place, left of = i2, label = below : $in_2$, fill = mgreen!60] (in2) {};
            \node[place, above right of = t1, label = below : $p_2$, fill = golden!60] (p2) {};
            \node[transition, right of = p2, label = below : $o_1$, fill = golden!60] (o1) {};
            \node[place, right of = o1, label = below : $out_1$, fill = golden!60] (out1) {};
            \node[place, below right of = t1, label = below : $p_4$, fill = mgreen!60] (p4) {};
            \node[transition, right of = p4, label = below : $o_2$, fill = mgreen!60] (o2) {};
            \node[place, right of = o2, label = below : $out_2$, fill = mgreen!60] (out2) {};
            \draw[double,->,mgreen!60] (p3) -- (t1);
            \draw[double,->,mgreen!60] (t1) -- (p4);
            \path[]
            (in1) edge[golden!60] (i1)
            (i1) edge[golden!60] (p1)
            (p1) edge[golden!60] (t1)
            (t1) edge[golden!60] (p2)
            (p2) edge[golden!60] (o1)
            (o1) edge[golden!60] (out1)
            (in2) edge[mgreen!60] (i2)
            (i2) edge[mgreen!60] (p3)
            (p4) edge[mgreen!60] (o2)
            (o2) edge[mgreen!60] (out2)
            (o2) edge (p2)
            ;
    \end{tikzpicture}}
    \vspace{.2cm}
    \caption{An OC WF-net with two object types $\otype_1$ (in yellow) and $\otype_2$ (in green)}
    \label{fig:example-2-1}
\end{figure}

Let us consider an example of \emph{tracking extension}.
Figure~\ref{fig:example-2-1} shows a \wfonet $N$ consisting of two nets related to two object types $\otype_1$ and $\otype_2$ that, for simplicity, are respectively depicted with green and yellow colors.

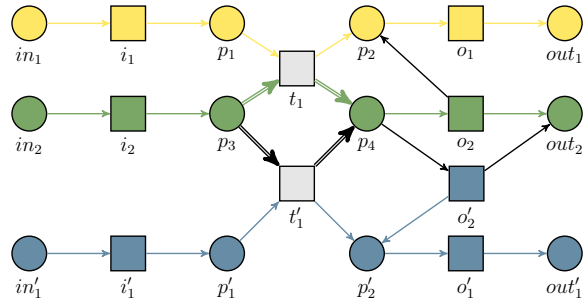
\begin{figure}[h!]
    \centering
    \resizebox{.63\textwidth}{!}{\begin{tikzpicture}[->,>=stealth',auto,x=1cm,y=.8cm,node distance=20mm and 8mm,thick, font=\large]
            
            \node[place, label = below : $in_1$, fill = golden!60] (in1) {};
            \node[transition, right of = in1, label = below : $i_1$, fill = golden!60] (i1) {};
            \node[place, right of = i1, label = below : $p_1$, fill = golden!60] (p1) {};
            \node[transition, below right of = p1, yshift=5mm,label = below : $t_1$] (t1) {};
            \node[place, below left of = t1, yshift=5mm, label = below : $p_3$, fill = mgreen!60] (p3) {};
            \node[transition, left of = p3, label = below : $i_2$, fill = mgreen!60] (i2) {};
            \node[place, left of = i2, label = below : $in_2$, fill = mgreen!60] (in2) {};
            \node[place, above right of = t1, yshift=-5mm, label = below : $p_2$, fill = golden!60] (p2) {};
            \node[transition, right of = p2, label = below : $o_1$, fill = golden!60] (o1) {};
            \node[place, right of = o1, label = below : $out_1$, fill = golden!60] (out1) {};
            \node[place, below right of = t1, yshift=5mm, label = below : $p_4$, fill = mgreen!60] (p4) {};
            \node[transition, right of = p4, label = below : $o_2$, fill = mgreen!60] (o2) {};
            \node[place, right of = o2, label = below : $out_2$, fill = mgreen!60] (out2) {};
            \node[transition, below right of = p3, label = below : $t_1^{\prime}$] (t1prime) {};
            \node[place, below left of = t1prime, label = below : $p_1^{\prime}$, fill = indigo!60] (p1prime) {};
            \node[transition, left of = p1prime, label = below : $i_1^{\prime}$, fill = indigo!60] (i1prime) {};
            \node[place, left of = i1prime, label = below : $in_1^{\prime}$, fill = indigo!60] (in1prime) {};
            \node[place, below right of = t1prime, label = below : $p_2^{\prime}$, fill = indigo!60] (p2prime) {};
            \node[transition, right of = p2prime, label = below : $o_1^{\prime}$, fill = indigo!60] (o1prime) {};
            \node[place, right of = o1prime, label = below : $out_1^{\prime}$, fill = indigo!60] (out1prime) {};
            \node[transition, below right of = p4, xshift=6mm, label = below : $o_2^{\prime}$, fill = indigo!60] (o2prime) {};
            \draw[double,->,mgreen!60] (p3) -- (t1);
            \draw[double,->,mgreen!60] (t1) -- (p4);
            \draw[double,->] (p3) -- (t1prime);
            \draw[double,->] (t1prime) -- (p4);
            \path[]
            (in1) edge[golden!60] (i1)
            (i1) edge[golden!60] (p1)
            (p1) edge[golden!60] (t1)
            (t1) edge[golden!60] (p2)
            (p2) edge[golden!60] (o1)
            (o1) edge[golden!60] (out1)
            (in2) edge[mgreen!60] (i2)
            (i2) edge[mgreen!60] (p3)
            (p4) edge[mgreen!60] (o2)
            (o2) edge[mgreen!60] (out2)
            (o2) edge (p2)
            (in1prime) edge[ indigo!60] (i1prime)
            (i1prime) edge[indigo!60] (p1prime)
            (p1prime) edge[ indigo!60] (t1prime)
            (t1prime) edge[indigo!60] (p2prime)
            (p2prime) edge[indigo!60] (o1prime)
            (o1prime) edge[indigo!60] (out1prime)
            (p4) edge (o2prime)
            (o2prime) edge (out2)
            (o2prime) edge[indigo!60] (p2prime)
            ;
    \end{tikzpicture}}
    \vspace{.2cm}
    \caption{Tracking 
		OC WF-net for an object of type $\otype_1$ }
    \label{fig:example-2-2}
\end{figure}

\begin{figure}[t!]
    \centering
    \resizebox{.63\textwidth}{!}{\begin{tikzpicture}[->,>=stealth',auto,x=1cm,y=.8cm,node distance=20mm and 8mm,thick, font=\large]
            
            \node[place, label = below : $in_1$, fill = golden!60] (in1) {};
            \node[transition, right of = in1, label = below : $i_1$, fill = golden!60] (i1) {};
            \node[place, right of = i1, label = below : $p_1$, fill = golden!60] (p1) {};
            \node[transition, below right of = p1, label = below : $t_1$] (t1) {};
            \node[place, below left of = t1, label = above : $p_3$, fill = mgreen!60] (p3) {};
            \node[transition, left of = p3, label = below : $i_2$, fill = mgreen!60] (i2) {};
            \node[place, left of = i2, label = below : $in_2$, fill = mgreen!60] (in2) {};
            \node[place, above right of = t1, label = below : $p_2$, fill = golden!60] (p2) {};
            \node[transition, right of = p2, label = below : $o_1$, fill = golden!60] (o1) {};
            \node[place, right of = o1, label = below : $out_1$, fill = golden!60] (out1) {};
            \node[place, below right of = t1, label = above : $p_4$, fill = mgreen!60] (p4) {};
            \node[transition, right of = p4, label = below : $o_2$, fill = mgreen!60] (o2) {};
            \node[place, right of = o2, label = below : $out_2$, fill = mgreen!60] (out2) {};
            \node[transition, below right of = p3, label = below : $t_1^{\prime\prime}$] (t1dprime) {};
            \node[place, below left of = t1dprime, label = above : $p_3^{\prime}$, fill = red!80] (p3prime) {};
            \node[transition, left of = p3prime, label = below : $i_2^{\prime}$, fill = red!80] (i2prime) {};
            \node[place, left of = i2prime, label = below : $in_2^{\prime}$, fill = red!80] (in2prime) {};
            \node[place, below right of = t1dprime, label = above : $p_4^{\prime}$, fill = red!80] (p4prime) {};
            \node[transition, right of = p4prime, label = below : $o_2^{\prime}$, fill = red!80] (o2prime) {};
            \node[place, right of = o2prime, label = below : $out_2^{\prime}$, fill = red!80] (out2prime) {};
            \node[transition, below right of = p3prime, label = below : $t_1^{\prime}$] (t1prime) {};
            \draw[double,->,mgreen!60] (p3) -- (t1);
            \draw[double,->,mgreen!60] (t1) -- (p4);
            \draw[double,->] (p3) -- (t1dprime);
            \draw[double,->] (t1dprime) -- (p4);
            \path[]
            (in1) edge[golden!60] (i1)
            (i1) edge[golden!60] (p1)
            (p1) edge[golden!60] (t1)
            (t1) edge[golden!60] (p2)
            (p2) edge[golden!60] (o1)
            (o1) edge[golden!60] (out1)
            (in2) edge[mgreen!60] (i2)
            (i2) edge[mgreen!60] (p3)
            (p4) edge[mgreen!60] (o2)
            (o2) edge[mgreen!60] (out2)
            (o2) edge[bend right =30] (p2)
            (in2prime) edge[red!80] (i2prime)
            (i2prime) edge[red!80] (p3prime)
            (p3prime) edge[red!80] (t1dprime)
            (t1dprime) edge[red!80] (p4prime)
            (p4prime) edge[red!80] (o2prime)
            (o2prime) edge[red!80] (out2prime)
            (o2prime) edge[bend right =50] (p2)
            (p1) edge[bend right =70] (t1dprime)
            (t1dprime) edge[bend right =70] (p2)
            (p1) edge[bend right =60] (t1prime)
            (t1prime) edge[bend right =60] (p2)
            (p3prime) edge[red!80] (t1prime)
            (t1prime) edge[red!80] (p4prime)
            ;
    \end{tikzpicture}}
    \vspace{.2cm}
    \caption{Tracking OC WF-net for an object of type $\otype_2$}
    \label{fig:example-2-3}
\end{figure}
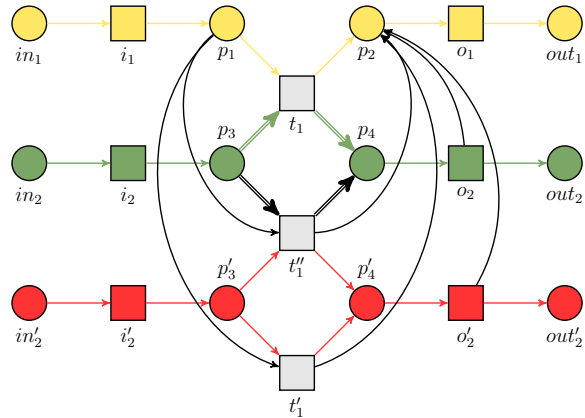

Let us consider tracking extensions for both object types of $N$.  
By properly identifying the tracking component for $\otype_1$, we demonstrate $N$'s $(\otype_1,\otype_1')$-tracking extension in Figure~\ref{fig:example-2-2} (here, $\otype_1'$  is represented with indigo color). 
Similarly to $\otype_1$, we present $N$'s $(\otype_2,\otype_2')$-tracking extension in Figure~\ref{fig:example-2-3} ($\otype_2'$ is in red).

%% file: 4-soundness.tex
\section{Object-centric Soundness}
\label{sec:soundness}
The most crucial correctness criterion for workflow nets is soundness \cite{AHHS11}. 
For classical WF-nets, a workflow case execution terminates properly, iff 
starting at the initial marking with a single token in its source place it terminates with a single token in its sink place, i.e. there are no ``garbage'' tokens after  termination. A  WF-net is called sound iff its execution  can terminate properly, starting at any reachable marking.

The soundness definition from above always considers one object (or, more traditionally, one case) in isolation, or several objects of the same type executing a common task, so that they start and finish together (generalized soundness). 
At the same time, as Definition~\ref{def:wfonet} suggests, \wfonets need to account for multiple objects of different types, when objects of the same type follow similar tasks, defined by a common for them workflow, and each object needs to terminate properly. The motivating example in Section~\ref{ex:students-projects} illustrates such a system.
To this end, we extend the classical notion of soundness with object types as well as multiple typed source and sink places.

\bigskip

We start by recalling the classical notion of soundness used for WF-nets and then show how it can be extended to the object-centric case. 
	\begin{definition}[WF-net soundness]
		Let $\net=(P,T,F)$ be a WF-net, and let $\inp, \outp\in P$ be its source  and sink places, respectively. 
		The net is called \emph{sound} iff
		\begin{enumerate}
			\item for every $\marking\in\reachable{\net}{[\inp]}$, it holds that $[\outp]\in\reachable{\net}{\marking}$ (\emph{option to complete});
			\item for every $\marking\in\reachable{\net}{[\inp]}$, if $[\outp]\subseteq\marking$, then $\marking=[\outp]$ (\emph{proper completion});
			\item for every $t\in\transitions$, there exists $\marking\in\reachable{\net}{[\inp]}$, s.t. $\enabled{\marking}{t}$ (\emph{no dead transitions}).
		\end{enumerate}
	\end{definition}

Now we come to defining object-oriented soundness for \wfonets. 

\begin{definition}[Object soundness, system soundness]
Let $\objnet$ be a $\otypes$-typed \wfonet, $\otype\in\otypes$ be one of its types, $\otype'\not\in\otypes$, and $\objnet_{\otype+\otype'}=(\otypes', \places',\transitions', \flow',\typing',\ell',A)$ be the  $(\otype, \otype')$-typed tracking extension of $\objnet$. 

$\objnet$ is called \emph{object-sound for type $\otype$}  (\emph{$\otype$-sound} for short) iff 
for every input marking $\marking_\inp\in\mult{{\places'}_\inp}$ in $\objnet_{\otype+\otype'}$ s.t. $\marking_\inp(\inp_\otype)=1$,
it holds that for any marking $M\in\reachable{\objnet}{\marking_\inp}$, there exist initial marking $\marking_\inp'\in\mult{(\places'_\inp)}$ and target marking $\marking'\in\mult{(\places')}$ s.t. :
\begin{itemize}
	\item $\marking'\in\reachable{\objnet'}{\marking + \marking_\inp'}$, i.e., target marking $M'$ can be reached from $M$, possibly   by adding  more tokens to source places;
	\item $\marking'(\outp_\otype) = 1$, i.e., the (tracked) object of type $\otype$ has reached its final state in target marking $\marking'$;
	\item for all $p\in \places'$ s.t. $\typing(p)=\otype$ and $p\not = \inp_\otype$, we have $\marking'(p) = 0$, i.e., the (tracked) object of type $\otype$ terminates properly (without leaving ``garbage''  behind, i.e. in the tracking component),
	\item 
	there are no dead transitions in  1-safe $\otype$-typed projection $\projection{\objnet}{\otype}^1$  of  \onet $\objnet$ with the initial marking that contains the only token in source place $\inp_\otype$.
\end{itemize} 
The same $\objnet$ is called \emph{object-centric sound} (\emph{OC-sound}, for short)  iff it is $\otype$-sound for every $\otype\in\otypes$.
\end{definition}

As in classical WF-nets, this definition  focuses on a ``concrete'' object and traces its lifecycle until the end. 
This is possible due to the relaxed treatment of object identifiers: the \wfonet cannot inspect the content of its tokens, but 
can always check types of places that carry the tokens. 
Thus, it suffices to look into quantities of objects available in a net marking.

However, unlike in classical WF-nets, this single object is not considered in isolation. 
Our notion of OC-soundness allows to account for potentially multiple objects in the source places, 
but cares only for one single object when it comes to checking whether the process has been completed or not 
(that is, we want to check only whether the ``object of interest'' has reached the source place). 
At the same time, we take no notice of other objects that were potentially ``helping'' that very selected object to complete the process. 
This means that the net may still contain tokens corresponding to other objects in some intermediate places.

\begin{example}
Consider the scenario discussed in Section~\ref{ex:students-projects}. There, a project can always be completed regardless ``concrete'' team members and leaders that are currently present in the net (i.e., we can always introduce into the net marking additional students to complete ``the selected'' project). 
Likewise, each team member has an opportunity to complete their task.
\end{example}

The following statements follow directly from the definition of OC-soundness.

\begin{proposition}
Let $\net=(\places,\transitions,\flow)$ be a WF-net. If $\net$ is sound, then it is also OC-sound as \wfonet $\objnet=(\set{\otype},\places,\transitions,\flow,\typing)$, where for all $p\in \places$, $\typing(p) =\otype$. 
\end{proposition}

Let us now come back to the projections. 
We now show how \wfonets and their 1-safe projections can be related.

\begin{proposition}
Let $\objnet=(\otypes, \places,\transitions,\flow,\typing,\ell,A)$ be an \wfonet. If $\objnet$ is OC-sound, 
then for any $\otype\in\otypes$, its 1-safe $\otype$-typed projection $\projection{\objnet}{\otype}^1$ is sound as a WF-net.
\end{proposition}

From this proposition it follows that the reachability set (set of all reachable markings) of a 1-safe projection $\projection{\objnet}{\otype}^1$ of sound \wfonet $\objnet$ is finite, since the reachability set of sound (classical) WF-net is always finite.

\begin{proposition}
Each 1-safe projections of an OC-sound \wfonet has finitely many reachable markings.
\end{proposition}

\medskip

For \wfonets we do not fix either input, or output markings. Sound \wfonets have, generally speaking, an infinite number of reachable markings. When defining OC-soundness, we take care of the possibility to  terminate properly for one selected object. For this object, all other objects play the role of resources required to complete its task. Unlike to workflow nets with resources, in \wfonets, these resources are dynamic and can be added as needed. A well-designed system should guarantee ``equal rights'' to all participants, i.e. each object has an ability to terminate properly  from any reachable marking. OC-soundness defines this formally. The good news is that this crucial property is decidable.

\begin{theorem}
The problem of checking OC-soundness for  an \wfonet is decidable.
\end{theorem}

%% file: 5-related.tex
\section{Related Work}\label{s:related}

\subsection{Modeling Object-centric and Data-aware Processes using Petri nets}

There are multiple Petri net-based approaches used for modeling data- and object-aware processes.
Here we briefly cover only the most recent ones.

The formalism of Petri nets with identifiers (PNIDs) was studied in~\cite{HSVW09}. In such nets, tokens carry tuples of object identifiers (or references). Moreover, transitions can produce globally new data values that are different from those in a current marking. PNIDs have been also used as the process modeling language in the Information Systems Modeling Language (ISML) presented in~\cite{PolyvyanyyWOB19}. In ISML, PNIDs are paired with special CRUD operations defining operations over relevant data. Such data are structured according to an ORM model.

A viewpoint on multi-case and data-aware process models is covered by works on so-called \emph{Proclets}.
This concept was first proposed in~\cite{AalstBEW00}.
Proclets are essentially quasi-workflows representing lifecycles of particular objects and allowing for different types of communication between such lifecycles via dedicated channels.
The Proclet notation was applied to model processes with complex interactions between process objects \cite{Fahland19} as well as artifact-centric processes \cite{Fahland193}.
Process artifacts here are documents, orders, deliveries, and other types of data objects which are manipulated within the process.
Artifact-centric process models consider these objects as first-class citizens.
Note that in such models process cases are often related to object types that leads to inter-case dependencies when more complicated communication patterns are involved.

In~\cite{GhilardiGMR20} the authors introduced catalog-nets (CLog-nets). Conceptually, the formalism marries read-only relational databases with a variant of colored Petri nets~\cite{JeK09} which allows for generation of globally fresh data values upon transition firing. In CLog-nets, transitions are equipped with guards that simultaneously inspect the content of tokens and query facts stored in the database.

\subsection{Soundness for Workflow Nets and their Extensions}

The notion of soundness for classical WF-nets was initially discussed in~\cite{Aalst98}.\footnote{This notion coincides with the one presented in the beginning of Section~\ref{sec:soundness}.}
Since then, many different variants of soundness have been proposed for WF-nets and their non-high-level extensions. An extensive study on such variants as well as related decidability results can be found in~\cite{AHHS11}.
In the following, we briefly touch upon some of the soundness notions.
Perhaps one of the most important generalizations of classical soundness are $k$-soundness and generalized soundness \cite{HeeSV04}. 
A WF-net is $k$-
sound if, starting with $k$ tokens in its source place, it always properly terminates with $k$ tokens in its sink place. Hence this notion allows handling multiple individual processes (cases) in a WF-net. The classical soundness then coincides with 1-soundness. Generalized soundness is $k$-soundness for all $k$. It was proved in \cite{HeeSV04} that both $k$-soundness and generalized soundness are decidable.

One of the important aspects of workflow development concerns resource management.
In classical Petri nets, one can distinguish places, corresponding to control flow states, and places, representing different kinds of resources. 
There are various works that study different resource-aware extensions of WF-nets together with related notions of soundness.
In \cite{BBS07} the authors studied a specific class of WFR-nets for which soundness was shown to be decidable. 
In \cite{HeeSSV05,SidorovaS13} a more general class of Resource-Constrained Workflow Nets (RCWF-nets) was defined. 
The constraints are imposed on resources and require that all resources that are initially available are present again after all cases terminate, and that for any reachable marking, the number of available resources does not override the number of initially available resources.
In \cite{HeeSSV05} it was proven that for RCWF-nets with a single resource type generalized soundness can be effectively checked in polynomial time. 
Decidability of generalized soundness for RCWF-nets with an arbitrary number of resource places was shown in~\cite{SidorovaS13}.

All the above mentioned works assume resources to be permanent, i.e., they are
acquired and released later at the end of a (sub-)process. 
Since resources are never created, nor destroyed, the process state space is explicitly bounded.
To study a more general case of arbitrary resource transformations (that can
arise, for example, in open and/or adaptive workflow systems), 
in \cite{Bashkin2014} WF-nets with resources (RWF-nets) were defined. 
RWF-nets extended RCWF-nets from \cite{HeeSSV05} in such a way that resources can be generated or consumed during a process execution without any restrictions. 
Unfortunately, even 1-sound RWF-nets are not bounded in general, hence
existing soundness checking algorithms are no applicable in that case. 
The decidability of soundness for RFW-nets was shown for a restricted case, in which a RWF-net has a single unbounded resource place.

There are also other works that propose more high-level extensions of classical WF-nets as well as the related notion(s) of soundness. 
In~\cite{LeoniFM18,FelliLM19,LeoniFM20}, the authors investigated data-aware soundness for data Petri nets, in which a net is extended with guards manipulating a finite set of variables associated with the net. 
That type of soundness was shown to be decidable. 
In~\cite{MontaliR16}, the authors proposed both a workflow variant of $\nu$-Petri nets as well as its resource-aware extension. The authors also defined a suitable notion of soundness for such nets and demonstrated that it is decidable by reducing the soundness checking problem to a verification task over another first order logic-based formalism. 
\cite{HaarmannW20,CaseModels2020} considered the soundness property for BPMN process models with data objects that can be related to multiple cases.
The approach consists of several transformation steps: from BPMN to a colored Petri net and then to a resource-constrained workflow net. The authors then check k-soundness against the latter.

%% file: 6-conclusions.tex
\section{Conclusions}

In this work, we investigated a notion of soundness for a workflow variant of the recently proposed formalism of object-centric Petri nets, in which each object type is associated with a dedicated workflow net.
Given this structural characteristic, proposed soundness focuses only on local termination of a single instance of an object type and assumes that the entire net can have any number of tokens being assigned to other dedicated workflows. 
We also conjectured that checking this new variant of soundness is decidable. 
